\documentclass{emulateapj}

\usepackage{xfrac}
\usepackage{amsmath}

\usepackage[textsize=tiny]{todonotes}
\renewcommand{\vec}{\mathbf}
\newcommand{\Turbo}{{\scshape Turbo} }
\newcommand{\Mmag}{M_{\textrm{mag}}}
\newcommand{\Dmm}{D_{\mu\mu}}
\newcommand{\Delmm}{\Delta_{\mu\mu}}
\newcommand{\D}{\mathrm{d}}
\newcommand{\I}{\mathrm{i}}
\newcommand{\uu}{\mathbf u}
\newcommand{\bb}{\mathbf b}

\newcommand{\xx}{\mathbf x}
\newcommand{\vv}{\mathbf v}

\newcommand{\db}{\mathbf{\delta b}}
\newcommand{\dB}{\mathbf{\delta B}}
\newcommand{\BO}{\mathbf{B_0}}
\newcommand{\kk}{\mathbf k}
\newcommand{\kmin}{k_{\mathrm{min}}}
\newcommand{\EE}{\mathcal E}
\newcommand{\KK}{\mathcal K}
\newcommand{\eem}{\mathbf e_{\mathrm{mot}}}

\newcommand{\sigc}{\sigma_c}
\newcommand{\fit}{^{\mathrm{(fit)}}}
\newcommand{\RR}{\mathcal{R}}
\newcommand{\RRcorr}{\mathcal{R}_{\mathrm{corr}}}

\usepackage{graphicx}
\begin{document}
\title{Cosmic-ray pitch-angle scattering in imbalanced MHD turbulence simulations}
\author{Martin~S.~Weidl}
\affil{Max-Planck-Institut f\"ur Plasmaphysik, D-85748 Garching, Germany}
\author{Frank~Jenko}
\affil{Department of Physics and Astronomy, University of California, Los Angeles, CA 90095, USA}
\author{Bogdan Teaca}
\affil{Applied Mathematics Research Centre, Coventry University, Coventry CV1 5FB, United Kingdom}
\and
\author{Reinhard Schlickeiser}
\affil{Institut f\"ur Theoretische Physik, Lehrstuhl IV: Weltraum- und Astrophysik, Ruhr-Universit\"at Bochum, D-44780 Bochum, Germany}

\begin{abstract}
Pitch-angle scattering rates for cosmic-ray particles in magnetohydrodynamic (MHD) simulations with imbalanced turbulence are calculated for fully evolving electromagnetic turbulence. We compare with theoretical predictions derived from the quasilinear theory of cosmic-ray diffusion for an idealized slab spectrum and demonstrate how cross helicity affects the shape of the pitch-angle diffusion coefficient.  Additional simulations in evolving magnetic fields or static field configurations provide evidence that the scattering anisotropy in imbalanced turbulence is not primarily due to coherence with propagating Alfv\'en waves, but an effect of the spatial structure of electric fields in cross-helical MHD turbulence.
\end{abstract}


\section{Introduction}
\label{secIntro}

Both turbulence and magnetic fields are phenomena which permeate essentially every subdiscipline of astrophysics. Studying the combined effect of both phenomena on fundamental processes such as the diffusion of charged particles in the context of magnetohydrodynamics is therefore a task of utmost importance for gaining a better understanding of the universe. 

The magnetohydrodynamic (MHD) turbulence of many astrophysical systems exhibits high degrees of cross helicity, which means that fluctuations of the bulk velocity of the medium and of magnetic fields induced in the plasma are strongly correlated. This property, which is also known as Alfv\'enicity when referring to the solar wind \citep{matthaeus2004transport} or as imbalanced turbulence in the astrophysical community \citep{perez2009role}, has been measured in observations of the fast solar wind \citep{marsch1990radial} and is important upstream of supernova remnant (SNR) shocks \citep{schlickeiser2008cosmic} as well as a driving force of the dynamo effect in accretion disks and young galaxies \citep{yoshizawa1993turbulent,brandenburg1998magnetic}.

The rate at which MHD turbulence scatters the pitch-angle of charged particles is a crucial parameter in many of these settings, particularly for diffusive shock acceleration in SNR shock fronts or measurements of velocity statistics in the solar wind. For example, the isotropization of cosmic-ray particles at SNR shock fronts is often considered to occur at the rate of Bohm diffusion, which can only be strictly derived in the limit of strong turbulence \citep{shalchi2009diffusive}. In weak or intermediate magnetic turbulence with $\delta B \lesssim B_0$, isotropization will be significantly slower and the efficiency of the diffusive shock acceleration process will be diminished. 

{Pitch-angle scattering on short timescales is particularly important in quasiperpendicular shocks. It has been proposed \citep{sagdeev1966} that low-energy ions entering the shock foot from the upstream region can be trapped by the electrostatic potential drop in the ramp. Within a few gyroperiods, ions may be accelerated while surfing along the convective electric field parallel to the surface of the shock front. Thus they can gain even higher energies, which are required for subsequent efficient Fermi acceleration. The details of this process, however, depend on how fast pitch-angle scattering proceeds over these few gyroperiods \citep{kirk1989particle}.}

The acceleration of dust particles moving through the interstellar medium at velocities close to the Alfv\'en speed similarly depends on the rate of pitch-angle scattering \citep{lazarian2002grain, hoang2012revisiting}. Consequently, a large number of comparisons between analytical predictions of this rate and test-particle simulations has been published \citep{michalek1998cosmic,qin2009pitch,dalena2012magnetic,tautz2013pitch}.

Most of these studies assumed a power-law spectrum for the magnetic turbulence and neglected the influence of cross helicity. Starting with a random distribution of the complex phases of the individual wave modes, realizations of the magnetic field were generated from this analytic description and used to compute test-particle trajectories. While this scheme represents a computationally efficient method of creating turbulence with a wide inertial range, it is only useful for comparisons with a magnetostatic turbulence model as it fails to capture the nonlinear effects important for a realistic description of the dynamic evolution of the turbulence, and in particular of the electric fields \citep{arzner2006effect}.

These shortcomings can be avoided by using fields generated in numerical three-dimensional MHD simulations to compute the test-particle scattering \citep[e.~g.\,][]{lehe2009heating,lynn2012resonance}, at the cost of a narrower inertial range than is possible with the analytic approach. If one aims to investigate the effects of cross helicity, which manifest themselves primarily in the dynamic propagation of Alfv\'en waves and a reduction of the motional electric field $\eem=-\uu\times\bb$, a full-MHD approach becomes necessary to obtain realistic results at particle speeds comparable to the Alfv\'en velocity.

Only few authors have numerically investigated the scattering of cosmic-ray particles in plasmas of non-vanishing cross helicity. \citet{beresnyak2011numerical} studied the scattering of relativistic test-particles in balanced and imbalanced MHD fields, but did not report observing any influence of the cross helicity on the pitch-angle-averaged scattering frequency. \citet{teaca2014acceleration} found that the decreased magnitude of $\eem$ in cross-helical MHD turbulence slows down the isotropization of Alfv\'en-speed particles significantly.

Here we present our results on the pitch-angle dependence of test-particle scattering in simulations of cross-helical MHD turbulence and compare them to predictions of the pitch-angle diffusion coefficient  derived from quasilinear theory \citep{schlickeiser1989cosmic}. 

This article is structured as follows: After an overview of various turbulence models used in quasilinear theory and how they affect the predicted pitch-angle diffusion coefficient in imbalanced turbulence (Section~\ref{secTheory}), we discuss the computational methods employed in our investigation (Section~\ref{secMethods}), including a detailed analysis of the structure of electric and magnetic fields in MHD simulations of plasma turbulence with various cross helicities. Having explained the rationale behind our choice of diagnostics, we present the results on pitch-angle scattering for test-particle simulations in these cross-helical turbulent fields (Section~\ref{secPitch}) before we summarize our findings in order to draw several important conclusions (Section~\ref{secConclusions}).

\section{Quasilinear predictions}
\label{secTheory}

\subsection{Magnetostatic turbulence}
\label{subMStheory}
The diffusion of a beam of charged cosmic-ray particles with velocity $v$ in a turbulent magnetic field $\bb(\xx)$ is commonly described in terms of a Fokker-Planck equation for the cosmic-ray phase-space density $f(\mu,z)$, where $\mu = \cos (v_z/v)$ is the cosine of the pitch-angle and $z$ is the coordinate along the global mean magnetic field $B_0 \hat z$ \citep[e.~g.\,][]{schlickeiser2002cosmic,shalchi2009nonlinear}:
\begin{equation}
    \frac{\partial f}{\partial t} + v\:\mu\:\frac{\partial f}{\partial z} = \frac{\partial}{\partial \mu} \left(\Dmm\:\frac{\partial f}{\partial \mu} \right).
    \label{eqnFPE}
\end{equation}

The form of the quasilinear pitch-angle diffusion coefficient $\Dmm = \lim_{t\to\infty} \langle [\mu(t)- \mu(0)]^2 \rangle/2t$ was first derived for cosmic-ray particles in magnetostatic slab turbulence by \citet{jokipii1966cosmic}. Expanding the turbulent magnetic field $\bb(\xx) = B_0 \hat z + \delta\bb_\perp(z)$ in a perpendicular perturbation $\delta\bb_\perp(z) \perp \hat z$ that only depends on the coordinate along the zeroth-order mean field $B_0 \hat z$, he found that
\begin{equation}
    \Dmm = \frac{\pi}{4}\ \Omega^{2-s} (1-\mu^2)\ \frac{\dB^2}{B_0^{2}}\ |v \mu|^{s-1} (s-1) \kmin^{s-1}.
    \label{eqnDmm1}
\end{equation}
Here, the power spectrum of the magnetic field perturbations is assumed as a slab spectrum $|\delta\bb_\perp(k_z)|^2 \D k_z= \dB^2 (s-1) \kmin^{s-1} k_z^{-s} \D k_z$ with a spectral index $s$, a lower cutoff wavenumber at $\kmin$, and root-mean-square amplitude $\langle \delta\bb_\perp^2\rangle^{1/2} = |\dB|$.

As the typical velocity of cosmic-ray particles $v \sim c$ exceeds the typical Alfv\'en velocity in the interstellar medium $v_A \sim 10\,\mathrm{km\,s}^{-1}$ by several orders of magnitude, the magnetic turbulence can be approximated as being time-independent. In the limit of very small wave frequencies $\Omega \ll v \kmin$, we can also presume that the characteristic electric field strength $\delta E$ becomes negligibly small since Faraday's law implies $\delta E \sim \Omega |\dB|/k$. As a consequence of these assumptions, the cross helicity does not enter into equation~(\ref{eqnDmm1}).

\subsection{Electrodynamic turbulence}
\label{subEDtheory}
For charged particles with $v \sim v_A$, however, the propagation speed of shear-Alfv\'en waves must be considered.  As first shown by \citet{schlickeiser1989cosmic}, all powers of $\gamma = v_A/v$ must be included in the derivation of $\Dmm$ to obtain the correct result. Accounting for the time-dependence of the magnetic turbulence and, through Faraday's law, the resulting electric fields self-consistently, one finds for the quasilinear pitch-angle diffusion coefficient in incompressible MHD turbulence:
\begin{multline}
    \Dmm = \frac{\pi}{2}\ \Omega^{2}\ B_0^{-2}\ (1-\mu^2)\\
    \times\ \sum_{\beta=\pm1} |v \mu + \beta v_A|^{-1}\ \EE^\beta_B \left( k_{\mathrm{res}}^{(\beta)} \right)
    \label{eqnDmm2a}
\end{multline}

Here $\beta \in \{+1,-1\}$ distinguishes shear-Alfv\'en waves traveling along ($\beta=-1$) or opposite ($\beta=+1$) the mean-field axis, the only types of waves that we consider in our simplified model of incompressible slab turbulence. We neglect additional contributions to the Fokker-Planck equation proportional to $\partial_p f$ and $\partial_p^2 f$ that result in adiabatic focusing and stochastic heating, focusing only on the pitch-angle evolution of a monoenergetic particle distribution. 

The total turbulence energy of the system can be separated into the energies contained in counter- and co-propagating waves traveling opposite to or along the direction of the mean magnetic field, the so-called positive and negative Elsasser energies $\EE^\pm = \langle (\uu \pm \delta\bb)^2 \rangle/4$. In quasilinear theory, an analogous decomposition into two Alfv\'en-wave populations is used for only the magnetic turbulence energy as well \citep{dung1990influence}, and we have written the respective components of the magnetic-turbulence power spectrum as $\EE^\beta_B(k_z)$ above. The wave numbers resonant with each wave population are defined as $k_{\mathrm{res}}^{(\beta)} = \Omega/(v\mu+\beta v_A)$.  In case the power spectra for both wave populations are not equal, the degree of imbalance between both propagation directions can be captured in the normalized cross helicity
\begin{equation}
	\sigma_c = \frac{\EE^+ - \EE^-}{\EE^+ + \EE^-}.
        \label{eqnSigma}
\end{equation}

Written in terms of $\sigma_c$, in the electrodynamic case the pitch-angle diffusion coefficient $\Dmm$ becomes \citep{dung1990influence}
\begin{align}
    \Dmm &= \frac{\pi}{4} (1-\mu^2) \sum_{\beta=\pm} \Omega^{2-s^\beta} \frac{(\dB)^2}{B_0^{2}} (s^{\beta}-1) \kmin^{s^\beta-1} \nonumber\\
      &\times\ v^{s^\beta-1}\left( 1 + \beta \mu \gamma \right)^2\ |\mu + \beta \gamma|^{s^\beta-1}\ (1 + \beta \sigma_c).
    \label{eqnDmm2}
\end{align}
To obtain this equation, we assume that the magnetic turbulence spectrum can be decomposed into contributions by co- and counter-propagating Alfv\'en waves in the same manner as the Elsasser energies, so that the two spectra
\begin{equation}
    |\db^\pm(k_z)|^2 \D k_z= (\dB^{\pm})^2\ (1-s^\pm)\ \kmin^{s^\pm-1}\ k_z^{-s^\pm}\ \D k_z
\end{equation}
describe the magnetic turbulence due to the two Alfv\'en-wave populations, where $(\dB^\pm)^2 = \dB^2 (1\pm\sigma_c)$ is the root-mean-square of the magnetic turbulence in the two wave populations.

\subsection{Magnetodynamic turbulence}
\label{subMDtheory}

In order to distinguish between the pitch-angle scattering caused by the electric-field acceleration and that due to the magnetic field alone, we have performed test-particle runs in which the electric-field component of the Lorentz force was not included. Without electric heating, the pitch-angle diffusion equation~(\ref{eqnFPE}) becomes exact again, although $\Dmm$ must be adapted to this `magnetodynamic' turbulence model. Generalizing the results that \citet{shalchi2009analytical} published for balanced turbulence to arbitrary cross helicity, we obtain for the pitch-angle diffusion coefficient:
\begin{multline}
    \Dmm = \frac{\pi}{4}\ \sum_{\beta=\pm1} \Omega^{2-s^\beta} \frac{(\dB)^2}{B_0^{2}}\ (s^\beta-1)\ \kmin^{s^\beta-1}\ v^{s^\beta-1}\\
     \times (1-\mu^2)\ |\mu + \beta \gamma|^{s^\beta-1}\ (1 + \beta \sigma_c).
    \label{eqnDmm3}
\end{multline}

\begin{figure}[ht]
    \centering
    \includegraphics{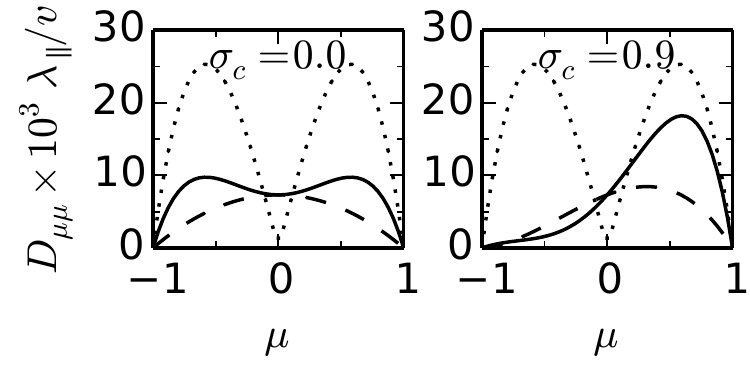}
    \caption{Shapes of the pitch-angle diffusion coefficient $\Dmm$ according to quasilinear theory for Alfv\'en-speed particles $(v=v_A)$ in magnetostatic turbulence (dotted), magnetodynamic turbulence (dashed), and fully electrodynamic MHD turbulence (solid lines) with amplitude $\delta B/B_0 = 0.1$ and Kolmogorov-like spectra $s^+ = s^- = 5/3$ for normalized cross helicities $\sigma_c = 0.0$ and $\sigma_c = 0.9$. ($\lambda_\parallel = 2\pi/\kmin$)}
    \label{figTheory}
\end{figure}

As one can easily check, eqns.~(\ref{eqnDmm2}) and (\ref{eqnDmm3}) are identical to (\ref{eqnDmm1}) if the Alfv\'en velocity becomes negligibly small compared to the particle velocity ($v \gg v_A$ or equivalently $\gamma \ll 1$) and both wave populations exhibit the same spectral exponent ($s^+ = s^-$). The quasilinear predictions for the pitch-angle diffusion coefficient in all three different cases in zero cross-helicity turbulence and strongly imbalanced turbulence are compared in Figure~\ref{figTheory}.

\section{Methods}
\label{secMethods}
\subsection{MHD setup}
\label{subMHD}

In order to compare the validity of the quasilinear slab-turbulence results for the pitch-angle diffusion with numerical results, we use the pseudospectral MHD code \Turbo \citep{teaca2009energy} to obtain representations of turbulent fields in which we propagate test-particles. Employing Alfv\'enic units, \Turbo solves the equations of resistive incompressible MHD on a cubic grid with periodic boundaries:
\begin{align}
 \partial_t \vec u &= - (\vec u \cdot \nabla) \vec u + \left(\vec b \cdot \nabla\right) \vec b + \nu \;\nabla^2 \vec u {+ \vec f^u} - \nabla \tilde p, \\
 \partial_t \vec b &= - (\vec u \cdot \nabla) \vec b + \left(\vec b \cdot \nabla\right) \vec u + \eta\;\nabla^2 \vec b {+ \vec f^b}
    \label{eqnMHDeqns}
\end{align}
Here $\tilde p = p/\rho$ is a re-normalized pressure that ensures that the velocity field remains divergence-free, $\nu$ and $\eta$ are kinetic viscosity and magnetic diffusivity, respectively, while $\vec f_u$ and $\vec f_b$ are forcing fields. The magnetic field $\bb$ includes a constant guiding field $\BO$ applied along the $z$ direction.

The cross-helical forcing scheme we use decomposes the Fourier-transformed fields $\uu(\kk)$ and $\bb(\kk)$ into eigenmodes of the curl operator. These helical eigenmodes for wave vectors in the ellipsoidal shell $2.5 \leq k_\perp^2 + 10^2 k_z^2 \leq 3.5$ are then forced separately in such a way that the injected amounts of energy ($\EE$), kinetic helicity ($\mathcal H_{\mathrm{kin}}$), magnetic helicity ($\mathcal H_{\mathrm{mag}}$), and cross-helicity ($\KK = \langle\uu\cdot\db\rangle = \sigma_c \EE$) are constant. Accordingly, we define the perpendicular and parallel correlation lengths as $\lambda_\perp=2\pi/3$ and $\lambda_\parallel=10\lambda_\perp$. The $k$-space ellipsoid is shortened by a factor of 10 along the $k_z$ axis to allow for anisotropic forcing of the turbulent fields so that the turbulence can attain stable steady-states in the presence of a magnetic guide field. In particular, turbulent energy can be supplied to the system at the rate $\epsilon = \partial \EE/\partial t$ without affecting any of the helicities, or cross-helicity can be injected at a desired rate $\sigma = (\partial \KK / \partial t)/\epsilon$ to drive the system towards an imbalanced steady-state with $\KK \neq 0$. 

The fields used for the test-particle runs were obtained on $512^3$ grids with physical side lengths of $2\pi~\times~ 2\pi~\times~20\pi$ and periodic boundary conditions. This anisotropically shaped simulation domain allows the turbulence to reach a steady-state in the presence of a strong magnetic mean-field along the $z$ direction, $B_0 = 10\, |\dB|$, with $|\dB| = \langle (\bb-\BO)^2 \rangle^{1/2}$.

Viscosity and magnetic diffusivity are chosen to ensure that Kolmogorov's dissipation length $\ell_K = (\nu^{3}/\epsilon)^{\sfrac14}$ is resolved by the numerical grid in all directions. Since we use isotropic viscosity and diffusivity coefficients, energy dissipation remains an essentially isotropic process, while the physical size of the numerical domain is anisotropic. Kolmogorov's dissipation length therefore corresponds to fewer grid cells in the direction along the mean-field than perpendicular to it, and resolving it with the same accuracy in both directions is impossible. With our choice of dissipation coefficients, Kolmogorov's dissipation length $\ell_K$ fulfills $k_{\Delta_\perp}\ell_K \sim 12$ and $k_{\Delta_\parallel}\ell_K \sim 1.2$, where $k_{\Delta_\perp}=\pi/\Delta x$ and $k_{\Delta_\parallel}=\pi/\Delta z$ are the wavenumbers corresponding to the grid resolution in perpendicular and parallel direction, respectively.

Driving turbulence with an energy-injection rate $\epsilon=0.1$ and cross-helicity injection rate $\sigma=0.0$ leads to a balanced steady-state with $\sigma_c=\KK/\EE=0.0$, while changing the cross-helicity injection to $\sigma=0.8$ eventually yields a steady-state cross helicity of $\sigma_c=0.9$. The latter value is obtained from averaging the scalar product $\KK = \langle \uu \cdot \delta\bb \rangle$, which saturates at $\KK = 0.9\ \EE$ when the dissipation of energy and cross helicity matches their injection rates. Similarly, injecting positive or negative cross helicity at a lower rate ($\sigma=\pm0.5$) results in a steady-state configuration with $\sigma_c = \pm0.6$, respectively.

\begin{figure}[t]
    \centering
    \includegraphics[width=\columnwidth]{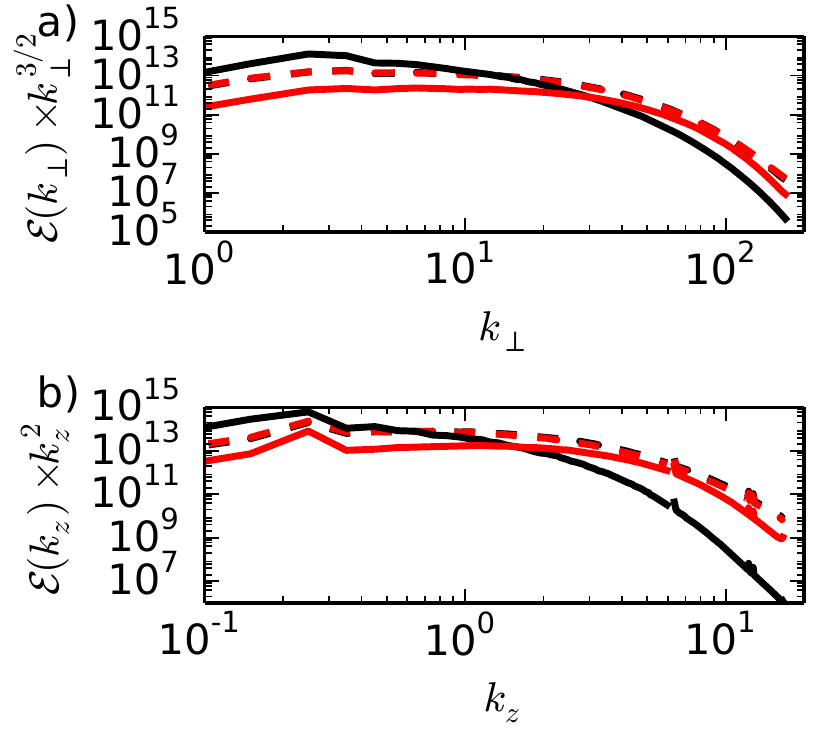}
    \caption{Steady-state spectra of positive (black) and negative (red) Elsasser energies, $\EE^\pm = (\uu \pm \db)^2/4$, for balanced ($\sigma_c=0.0$, dashed lines) and strongly imbalanced ($\sigma_c=0.9$, solid lines) cases. a)~Spectrum perpendicular to $\BO$. b)~Parallel spectrum}
    \label{figSpec}
\end{figure}

Figure~\ref{figSpec} provides comparisons of steady-state spectra for $\sigma_c = 0.0$ and $\sigma_c=0.9$. For the strongly imbalanced case (solid lines), the Elsasser energy with the same (in this case, positive) sign as the total cross-helicity dominates over the opposite Elsasser energy at low $k \lesssim 20$, while both Elsasser energies are in equipartition in the balanced case. The positive Elsasser energy is generally associated with waves propagating opposite to a strong magnetic mean-field. We deduce that the majority of the propagating shear Alfv\'en waves in the imbalanced case has a phase velocity directed opposite to the mean field (their energy is respresented by $\EE^+$), while the wave distribution in the balanced case exhibits no such anisotropy.

\begin{figure*}[ht]
    \centering
    \includegraphics[width=\textwidth]{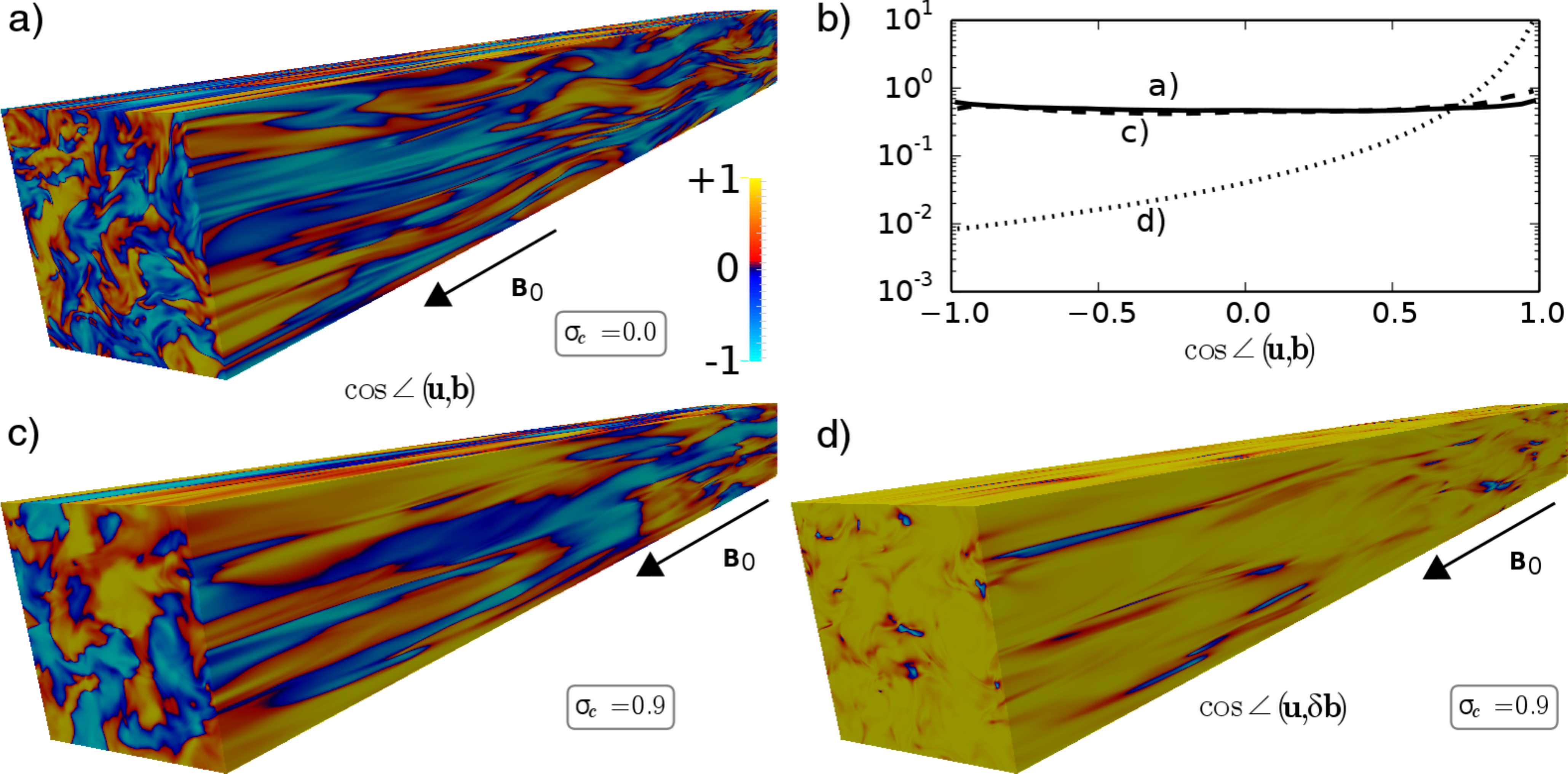}
    \caption{a) Alignment of velocity and magnetic fields, $\cos\angle(\uu,\bb)$, in balanced turbulence. b) Histograms of alignment in steady-state balanced turbulence (a), alignment of velocity and total magnetic field (c) and of velocity and magnetic fluctuations (d) in strongly cross-helical turbulence. c) Alignment distribution of $\cos\angle(\uu,\bb)$ in strongly cross-helical turbulence ($\sigma_c=0.9$). d) Distribution of $\cos\angle(\uu,\delta\bb)$ in the same snapshot of strongly cross-helical turbulence}
    \label{figCOSplots}
\end{figure*}

The spectra in the direction parallel to the mean field show a similar dominance of the Elsasser energy with the same sign as the cross-helicity (Figure~\ref{figSpec}). As analytical predictions and recent solar-wind measurements have shown (see \citet{podesta2009dependence} and references therein), one expects a scaling $\EE \propto k_z^{-2}$, and indeed this scaling fits our simulations well in the balanced case. In the cross-helical cases, the spectral indices of the positive and negative Elsasser energies differ slightly \citep{grappin1983dependence}. 

The imbalance of the cross-helical cases is also obvious in plots of the spatial distribution of the cosine between the velocity field and the magnetic field $\bb = \BO + \delta\bb$. The zero-cross-helicity case exhibits positive and negative values of $\cos\angle(\uu,\bb)$ in approximately equal proportion (Figure~\ref{figCOSplots}a,b). For the strongly cross-helical case with $\sigma_c = 0.9$, the velocity field shows a similarly weak correlation with the total magnetic field (Figure~\ref{figCOSplots}c; however, note the small changes in the slope of the histogram curve at the extremal values $\cos\angle(\uu,\bb)=\pm1$ in b), while positive values of $\cos\angle(\uu,\delta\bb)$ dominate if the alignment between the velocity field and only the fluctuations of the magnetic field is considered (Figure~\ref{figCOSplots}d). A similar plot of $\cos\angle(\uu,\delta\bb)$ for the balanced case is not depicted since it varies only negligibly from the total-field alignment already shown in Figure~\ref{figCOSplots}a.

Due to the almost identical distribution of the alignment of $\uu$ and $(\BO+\delta\bb)$ in balanced and cross-helical turbulence, both cases exhibit a similar level of isotropy of the motional electric field along the mean-field direction. The distribution of $e_z$ is symmetric around $e_z=0$ with an approximately exponential drop-off in both directions \citep{breech2003probability}, with the global average $\langle e_z \rangle = 0.000458 \langle | \mathbf e | \rangle$ in the balanced case and $\langle e_z \rangle = -0.000034 \langle | \mathbf e | \rangle$ in the strongly cross-helical case ($\sigc=0.9$).

\begin{figure}[htp]
    \centering
    \includegraphics[width=\columnwidth]{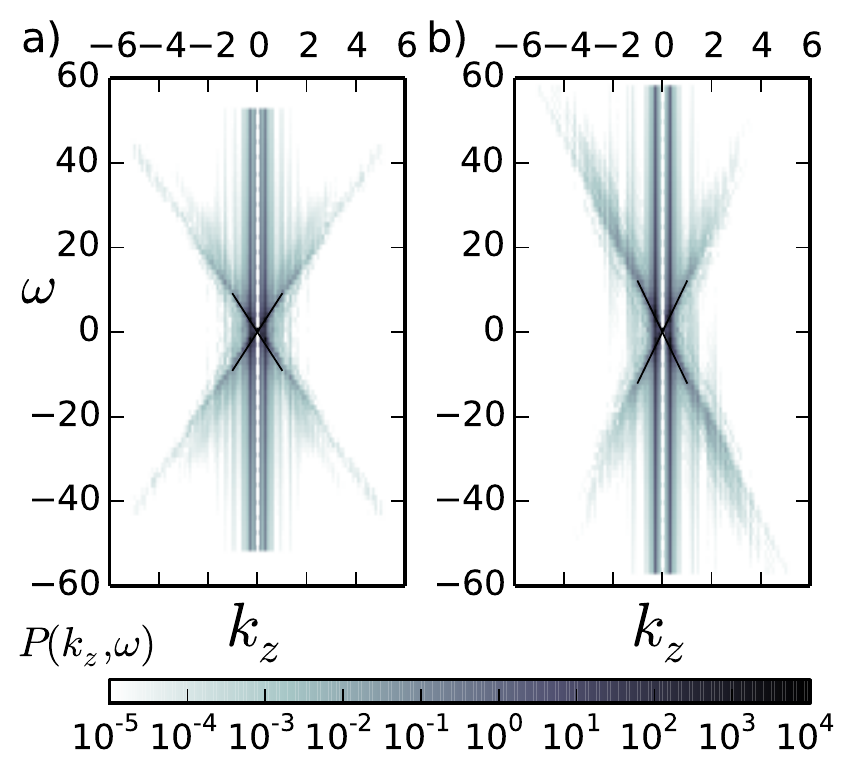}
    \caption{Power spectra $P(k_z,\omega)$ of a)~balanced ($\sigma_c=0.0$) and b)~strongly cross-helical ($\sigma_c=0.9$) MHD turbulence}
    \label{figDisp}
\end{figure}

In order to investigate the presence of incompressible Alfv\'en waves, we switched off forcing and dissipation once the balanced and imbalanced field configurations had reached a steady-state. Setting $\vec f_u = \vec f_b = \eta = \nu = 0$, we ensure that the propagation of shear-Alfv\'en waves is not disturbed by non-ideal effects, which are not accounted for in the derivation of the quasilinear diffusion coefficients. Comparing the power spectra of the magnetic field Fourier-transformed in space and time (Figure~\ref{figDisp}), we find evidence of Alfv\'en waves traveling along and opposite the mean-field direction and therefore confirm that the evolution of the MHD turbulence is shaped mainly by propagating shear-Alfv\'en waves. The $x$-shaped dispersion plot of the balanced case is symmetric with respect to the $k_z=0$ line, indicating that the waves propagating parallel and antiparallel to the mean field are equally strong. In the strongly cross-helical case ($\sigma_c=0.9$), there is more energy in the modes along $\omega = -v_A k_z$ than in the modes with the oppositely directed phase velocity along $\omega = +v_A k_z$, particularly at larger $k_z$, indicating once more a dominance of Alfv\'en waves propagating opposite to the mean- field.

\subsection{Test-particle setup}
\label{subTestparticle}

We continue using zero forcing and dissipation while we compute the forces that test-particles experience in these turbulent steady-state fields. The trajectories of these test-particles are evolved using an implicit fourth-order Runge-Kutta solver with an adaptively determined timestep size. Their velocity at each timestep is calculated using the Lorentz force computed from the MHD magnetic field $\bb$ and, in some runs, the motional electric field $\eem = - \uu \times \bb$, as interpolated from the grid data using third-order spline functions.

At the starting time $t=0$ of the test-particle simulation, ensembles of $N_p=2\times10^4$ particles are distributed randomly on the grid, with an energy equal to $E_p = v(0)^2/2$ and a velocity $v_z = \mu v(0)$ along the direction of the mean-field magnetic field. The initial pitch-angle cosine $\mu$ differs for each ensemble, taking values from -1.0 to 1.0. To demonstrate the interaction with propagating Alfv\'en waves, we initialize the test-particle ensembles with velocities close to the Alfv\'en velocity, using both $v(0)=v_A$ and $v(0)=3v_A$. The charge-to-mass ratio of the test-particles is chosen such that their initial gyroradius is $r_g(0)=v(0)/(qB_0/m)= 0.12 \lambda_\perp$ for Alfv\'en-speed particles and $r_g(0)=0.36\lambda_\perp$ for $v(0)=3v_A$.

Since our goal is to compare analytical pitch-angle diffusion coefficients predicted by quasilinear theory with simulation results, we define $\mu$ as the cosine of the angle subtended by the particle velocity and the direction of the global mean magnetic field (the $z$ direction) and thus follow the QLT convention. Using an only locally averaged mean-field direction as reference would yield a pitch-angle diffusion coefficient that would not fully conform to the quasilinear definition. QLT is derived by expanding the equations of motion in a small perturbation of a globally constant magnetic field, hence it is the global mean-field direction that is relevant for a comparison.

\subsection{Measuring pitch-angle scattering}
\label{subMeasuring}

In order to compare the pitch-angle scattering of test-particles on short timescales with quasilinear estimates, we define
\begin{equation}
    \Delmm(\mu(0)) = \frac12 \left\langle \frac{\left[ \mu(T_g) - \mu(0) \right]^2}{T_g} \right\rangle
    \label{eqnDelmm}
\end{equation}
as the scattering rate after one gyroperiod. Here $\mu(0)$ and $\mu(T_g)$ are the pitch-angle cosine at the test-particle injection time and after one gyroperiod, respectively. Whereas $\Dmm$, as defined above in the context of QLT, corresponds to an infinite-time limit, we decided to measure $\Delmm$ instead for several reasons. Although calculating $\Delmm$ for timescales significantly longer than $T_g$ could be expected to improve the agreement with quasilinear calculations of $\Dmm$, we will demonstrate that this is not the case, due to various effects which were not included in the derivation of the quasilinear coefficients above, but which are inevitably present both in our simulations and in realistic MHD turbulence. In Appendix~\ref{secAppendix}, we demonstrate that $\Delmm$ is bounded below by $\Dmm$.

First, as the pitch-angle cosine $\mu$ is limited to the interval $[-1,+1]$, its evolution is subject to boundary effects. Measuring the mean-square pitch-angle deviation
\begin{equation}
    \Delta\mu^2(t) = \left\langle \left[ \mu(t) - \mu(0) \right]^2 \right\rangle 
    \label{eqnDeltaMu}
\end{equation}
for large $t$ will, in general, underestimate the diffusion coefficient $\Dmm(\mu(0))$ at the initial $\mu(0)$ as some test-particles will have been reflected at the boundaries. Even before this reflection occurs, however, $\Dmm(\mu)$ varies strongly with $\mu$ and is therefore difficult to determine if the pitch-angle distribution spreads too quickly. This is particularly relevant if the turbulence amplitude is too large (for instance $\delta B/B_0 \sim 1$). In that case, even particles with a gyroradius as small as $r_g(0)=0.36\lambda_\perp$ isotropize almost completely within a single gyroperiod. Therefore, we only present results for an intermediate turbulence strength of $\delta B/B_0 = 0.1$. 

Second, our goal is to investigate the role of electric fields, which accelerate or decelerate test-particles on longer timescales. Since the Fokker-Planck equation~(\ref{eqnFPE}) does not account for these effects, good agreement can only be expected if the timescales on which $\Delmm$ is measured are short enough that the change in kinetic energy is negligible.

Most importantly, we will demonstrate that pitch-angle scattering on longer timescales is severely constrained by the approximate conservation of the magnetic moment,
\begin{equation}
    \Mmag = m \frac{v_\perp^2}{2B},
    \label{eqnMmag}
\end{equation}
the first adiabatic invariant in slowly varying magnetic fields \citep{landau1982mechanics}. The standard formulation of quasilinear theory does not account for the approximate conservation of $\Mmag$ (\citealp{goldstein1980mean,jaekel1998magnetic}; see however \citealp{schlickeiser2010cosmic}). As an adiabatic invariant, $\Mmag$ needs only be considered for the evolution of $\Delta\mu^2$ on timescales exceeding several gyroperiods, so that $\Delmm$, which is computed after only one gyroperiod, is not affected by these constraints and therefore allows for a better comparison with quasilinear predictions. Although the absolute value of $\Delmm$ will not necessarily be close to the value of $\Dmm$ in all cases, the scaling with the pitch-angle cosine $\mu$, which is our primary concern in these investigations of cross-helical turbulence, will be almost identical.

\section{Pitch-angle scattering in cross-helical turbulence}
\label{secPitch}
\subsection{Fully electrodynamic runs}
\label{subED}
\begin{figure}[htp]
\centering
\includegraphics[scale=1.0]{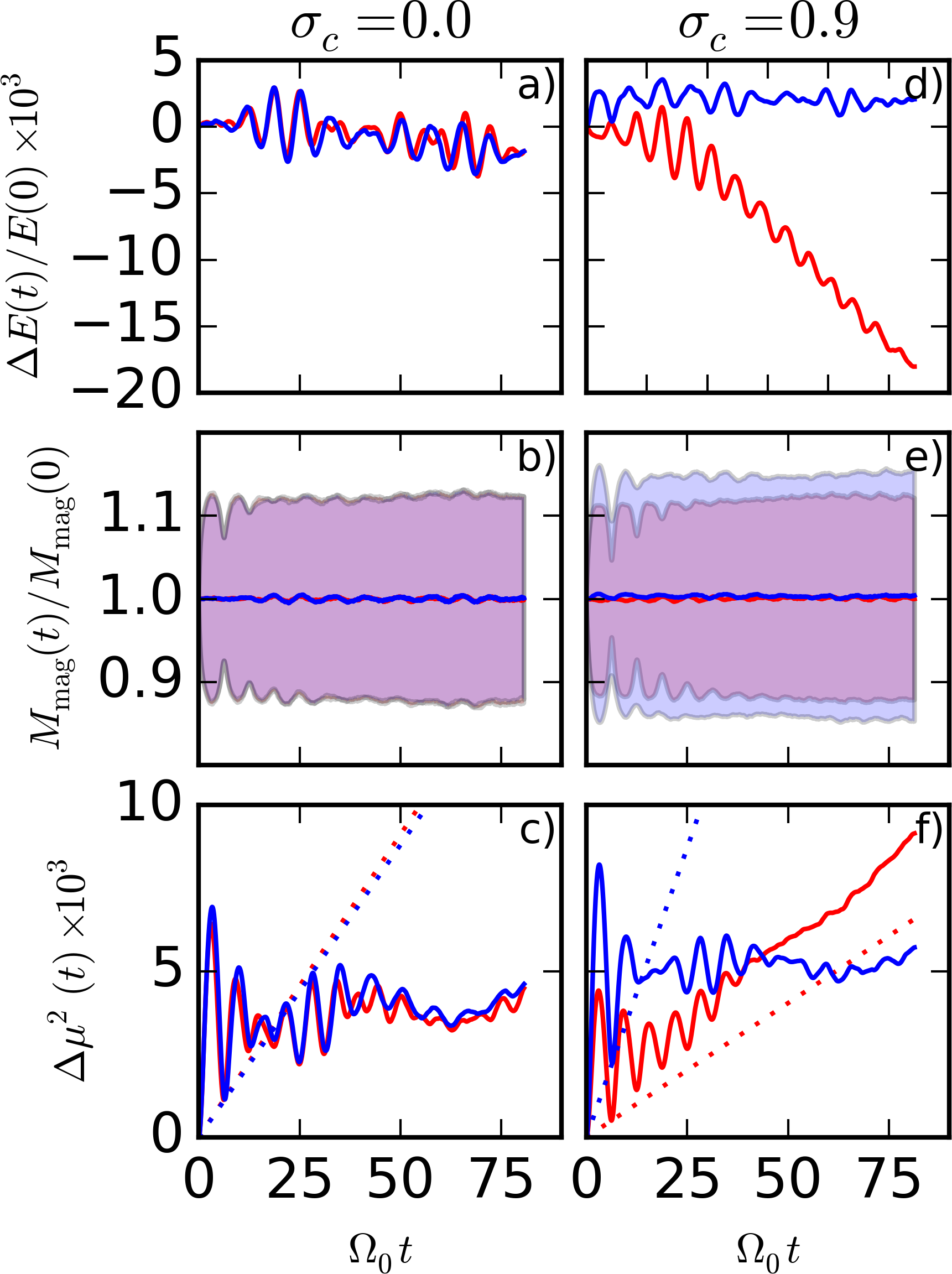}
\caption{Evolution in fully electrodynamic turbulence for the relative change of the kinetic energy $\Delta E$, the ensemble-averaged magnetic moment $\Mmag$ ($1\sigma$-deviation shaded in blue/red), and the mean-square displacement of the pitch-angle cosine, $\Delta\mu^2$, for particles with $v(0)=3v_A$ and $\mu(0)=-0.6$ (red) and $\mu(0)=+0.6$ (blue) }
\label{figEMMbxu}
\end{figure}

We begin with the most realistic scenario, fully electrodynamic turbulence, in which the evolution of the MHD fields is computed in parallel with the test-particle trajectories. The initial conditions correspond to the field configurations presented in Subsection~\ref{subMHD}. In Figure~\ref{figEMMbxu}, we show the evolution of the kinetic energy, the magnetic moment, and the mean-square pitch-angle deviation for two representative ensembles with $v(0)=3v_A$ and $\mu(0)=\pm0.6$ in balanced and strongly imbalanced turbulence. 

\begin{figure}[htp]
\centering
\includegraphics[scale=1.00]{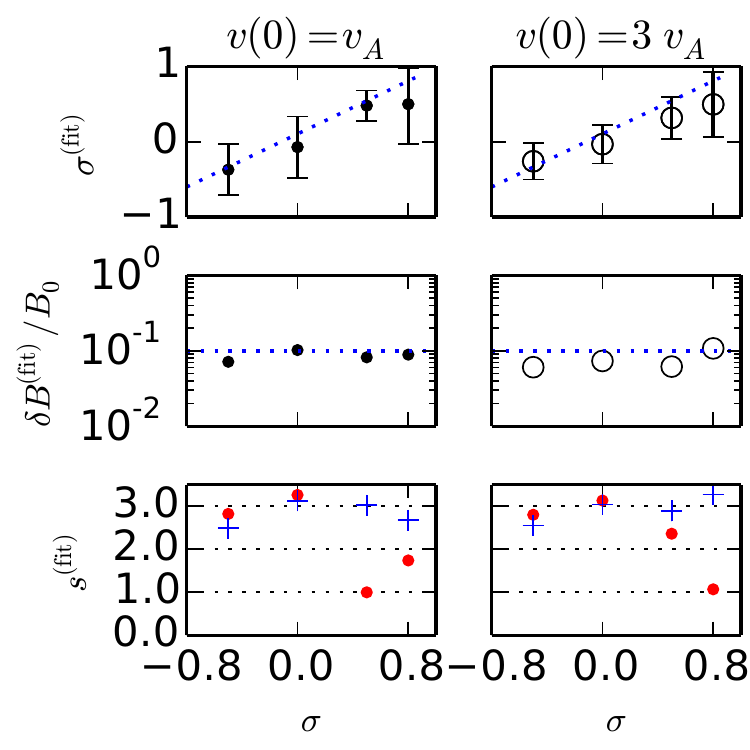}
\caption{Fully electrodynamic case: best-fit values with respect to equation~(\ref{eqnDmm2}) for the cross helicity $\sigma\fit$, the magnetic-turbulence amplitude $\delta B\fit$, and the spectral indices $s^+$ (blue crosses) and $s^-$ (red circles), for particles with $v(0)=v_A$ (left) and $v(0)=3v_A$ (right), for four values of $\sigma$}
\label{figEDsig}
\end{figure}

\begin{figure*}[htp]
    \centering
    \includegraphics{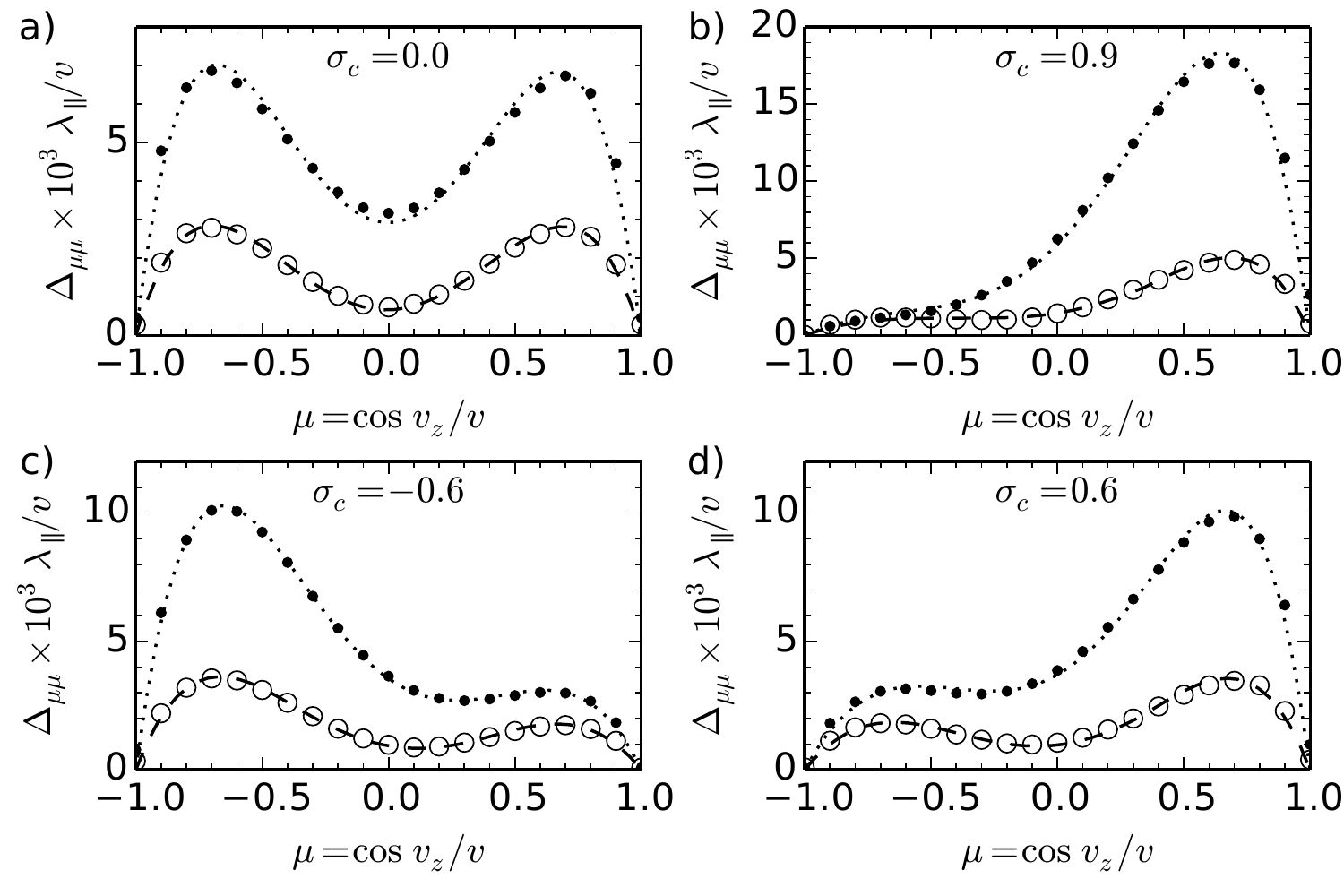}
    \caption{Fully electrodynamic case: pitch-angle scattering rate $\Delmm(\mu)$ for cosmic-ray particles with $v=v_A$ and gyroradius $r_0 = v/(\alpha B_0) = 0.12 \lambda_\perp$ determined numerically (solid dots) and best-fit curves from eqn.~(5) (dotted lines), and for particles with $v=3v_A$ and gyroradius $r_0 = 0.36 \lambda_\perp$ (hollow dots and dashed line). a)~Balanced turbulence with $\sigma_c=0.0$ b)~Strongly cross-helical turbulence with $\sigma_c=0.9$ c)~Cross-helical turbulence with $\sigma_c=-0.6$ d)~Cross-helical turbulence with $\sigma_c=0.6$}
    \label{figED}
    \vspace{0.5in}
\end{figure*}

For both ensembles in the balanced-turbulence case, energy and the averaged magnetic moment oscillate with the gyrofrequency, but the values of the minima attained after each full gyroperiod show only negligible variation. The mean-square deviation of $\Mmag$, as indicated by the shaded regions, converges to about 12 \% of the initial value as the magnetic moment is conserved to a large degree because of the relatively weak perturbation amplitude $\delta B/B_0$. This puts a limit to the evolution of $\Delta\mu^2(t)$, however, since the kinetic energy, the magnetic moment, and the pitch-angle cosine are approximately connected via 
\begin{equation}
    E \approx \Mmag B + E \mu^2.
    \label{eqnEMmagMu}
\end{equation}
This relation is valid if we assume that the magnetic turbulence is so weak that the difference between the averaged velocity components $\langle v_\parallel^2\rangle$ along the local magnetic-field direction and $\langle v_z^2 \rangle$ along the global mean-field can be neglected, the difference being of order $(\delta B/B_0)^2$. Consequently, the minima of $\Delta\mu^2(t)$ after each gyroperiod increase in value for two gyroperiods before settling on an almost constant value.

In imbalanced turbulence with $\sigc=0.9$, the test-particle ensemble with a positive pitch-angle of $\mu(0)=+0.6$ evolves similarly to the balanced case. However, the counter-propagating particles with $\mu(0)=-0.6$ are much closer to resonance with the dominating Alf\'en-wave population, which is traveling opposite to the mean-field direction as well. Consequently, these particles undergo deceleration as a result of inverse Landau damping and their kinetic energy exhibits a steady decrease (red lines in Figure~\ref{figEMMbxu}). While the magnetic moment is approximately conserved again, the variation in energy makes it possible for pitch-angle scattering to continue on longer timescales; hence the minima of $\Delta\mu^2(t)$ after each gyroperiod grow steadily. Obviously, a comparison of the mean-square pitch-angle deviations for co- and counter-propagating ensembles on these longer times would be misleading. Instead, as described above, we compare the values of the first minima of $\Delta\mu^2$ (or rather the slopes of the dotted lines through these minima).

These values of $\Delmm$ for 21 different initial pitch-angles ($\mu(0)\in\{-1,-0.9,\ldots,+1.0\}$) are then used to determine a best-fit parameter set $\{\delta B\fit, \sigma\fit, s^+, s^-\}$ for the quasilinear prediction for $\Dmm$ given by equation~(\ref{eqnDmm2}). The remaining parameters ($\kmin, \Omega_0, \gamma$) were fixed at their actual values before the optimization was initialized. Although equation~(\ref{eqnDmm2}) was derived under the assumption of a highly simplified slab spectrum with a constant power index, hardly what our simulations of realistic MHD exhibit in Figure~\ref{figSpec}, it will become clear that the agreement of $\Delmm$ and the quasilinear predictions of $\Dmm$ is rather striking. The results from a Levenberg-Marquardt fit using our measurements of $\Delmm(\mu)$ are shown in Figure~\ref{figEDsig}. 

The fit-values of $\sigma\fit$ match the values of $\sigma$ used for the cross-helicity injection well in all cases. A decrease of the fit quality if one Elsasser energy is very small (that is, for $\sigma=0.8$) is visible but not surprising. The relative turbulence amplitudes $\delta B\fit/B_0$ are close to the actual value $\delta B/B_0=0.1$, usually lying slightly below because the particle scattering is mainly due to only the slab component of the turbulence. The spectral indices obtained from the fit procedure are almost always larger than the predictions for the inertial-range spectrum in Kolmogorov's or Kraichnan's theories (5/3 or 3/2, respectively) since, like realistic turbulence, the spectra in our simulations steepen at the transition to the dissipation range (see Figure~\ref{figSpec}).

The graphs of the quasilinear $\Dmm$ for these parameter sets perfectly match the scattering rates $\Delmm$ determined from the test-particle trajectories, as shown in Figure~\ref{figED}. Whereas the graphs in the balanced case are symmetric with respect to the sign of $\mu(0)$, particles in the imbalanced cases are scattered faster within the first gyroperiod if they travel in the direction of the weaker Alfv\'en-wave population (if $\mu(0)>0$ for $\sigma>0$ and \textit{vice versa}).

\begin{figure}[tp]
\centering
\includegraphics{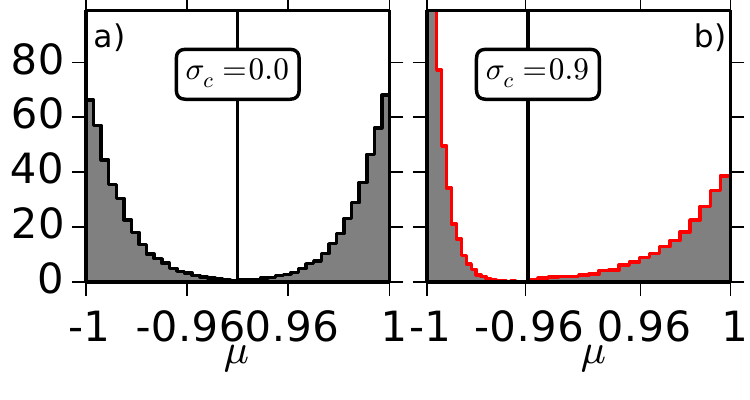}
\caption{Fully electrodynamic case: histograms of the pitch-angle cosine after ten gyroperiods for test-particles with $v(0)=3v_A$ and $\mu(0)=\pm1.0$ in a) balanced turbulence ($\sigma_c=0.0$), b) strongly imbalanced turbulence ($\sigma_c=0.9$)}
\label{figHisto2}
\end{figure}

This asymmetry is shown more directly in Figure~\ref{figHisto2}, which contains pitch-angle histograms of particles with $v(0)=3v_A$ after ten gyroperiods for test-particles initially propagating exactly parallel or anti-parallel to the mean-field direction. In this case, both the heating that was observed for the negative-$\mu$ ensemble in Figure~\ref{figEMMbxu}d and the initial magnetic moment are insignificantly small, and the behavior predicted by quasilinear theory can be observed for much longer. While both ensembles, with $\mu(0)=+1$ and with $\mu(0)=-1$, are scattered equally fast in balanced MHD turbulence, positive cross helicity leads to particles propagating along the mean-field direction being scattered much faster than in the opposite case.

\begin{figure}[tp]
\centering
\includegraphics{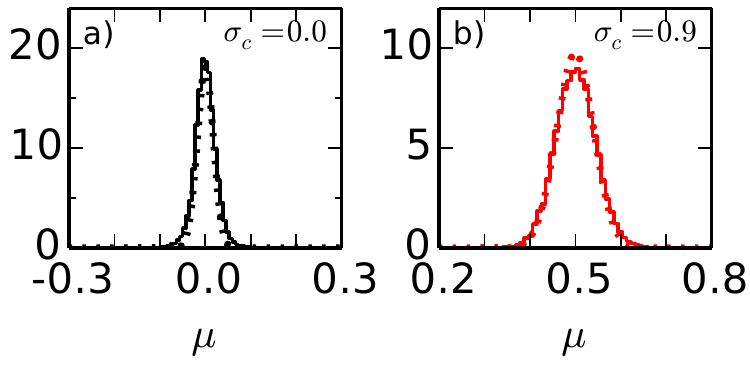}
\caption{Fully electrodynamic case: histograms of the pitch-angle cosine after one gyroperiod for test-particles with a) $v(0)=v_A$ and $\mu(0)=0.0$ in balanced turbulence with $\sigma_c=0.0$, b) $v(0)=3v_A$ and $\mu(0)=+0.5$ in strongly imbalanced turbulence with $\sigma_c=0.9$}
\label{figHistoFPE}
\end{figure}

In Figure~\ref{figHistoFPE}, we compare similar histograms of sample pitch-angle distributions after one gyroperiod with solutions that we directly obtained from the Fokker-Planck equation~(\ref{eqnFPE}), using $f(\mu,t=0) = \delta( \mu(0) -\mu)$ as initial condition and assuming a homogeneous density along the $z$-coordinate. While we showed above that the mean-square displacement of the pitch-angle cosine after one gyroperiod agrees with the scaling of $\Dmm(\mu)$ predicted by QLT, Figure~\ref{figHistoFPE} demonstrates that the shape of the histograms agrees with a diffusive broadening of the pitch-angle distribution within the first gyroperiod. This confirms once more that the quasilinear prediction~(\ref{eqnDmm2}) gives an accurate description of the pitch-angle scattering on short timescales even in evolving imbalanced turbulence with electric fields.

\subsection{Magnetodynamic runs}
\label{subMD}
There are two possible causes for this asymmetry in imbalanced turbulence: the coherent interaction with propagating Alfv\'en waves and the spatial structure of the turbulent fields. To identify the more important mechanism, we first ignore the electric-field component of the Lorentz force and propagate the test-particles using only
\begin{equation}
    \dot \vv_i = \alpha \vv_i \times \bb(\xx_i, t).
    \label{eqnLorentzMD}
\end{equation}

\begin{figure}[htp]
\centering
\includegraphics{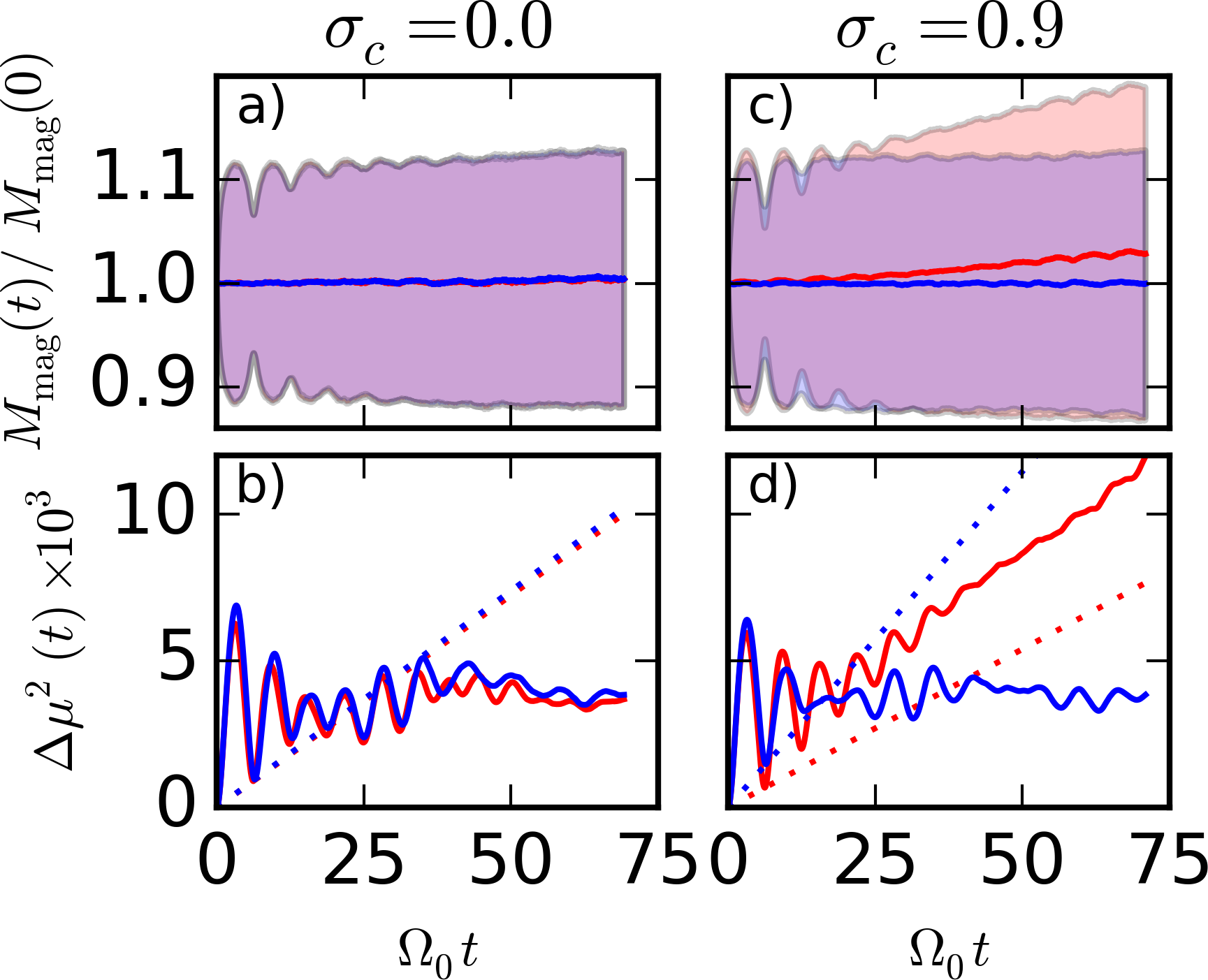}
\caption{Magnetodynamic case: evolution of $\Mmag$ and $\Delta\mu^2$ as in figure~\ref{figEMMbxu}}
\label{figEMMme}
\end{figure}

\begin{figure}[htp]
\centering
\includegraphics{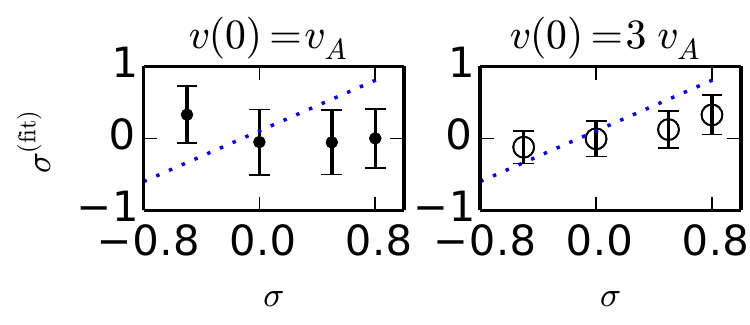}
\caption{Magnetodynamic case: best-fit values for equation~(\ref{eqnDmm3}) for $\sigma\fit$ for particles ensembles with $v(0)=v_A$ (left column) and $v(0)=3v_A$ (right), for four different values of $\sigma$}
\label{figMEsig}
\end{figure}

\begin{figure*}[ht]
    \centering
    \includegraphics{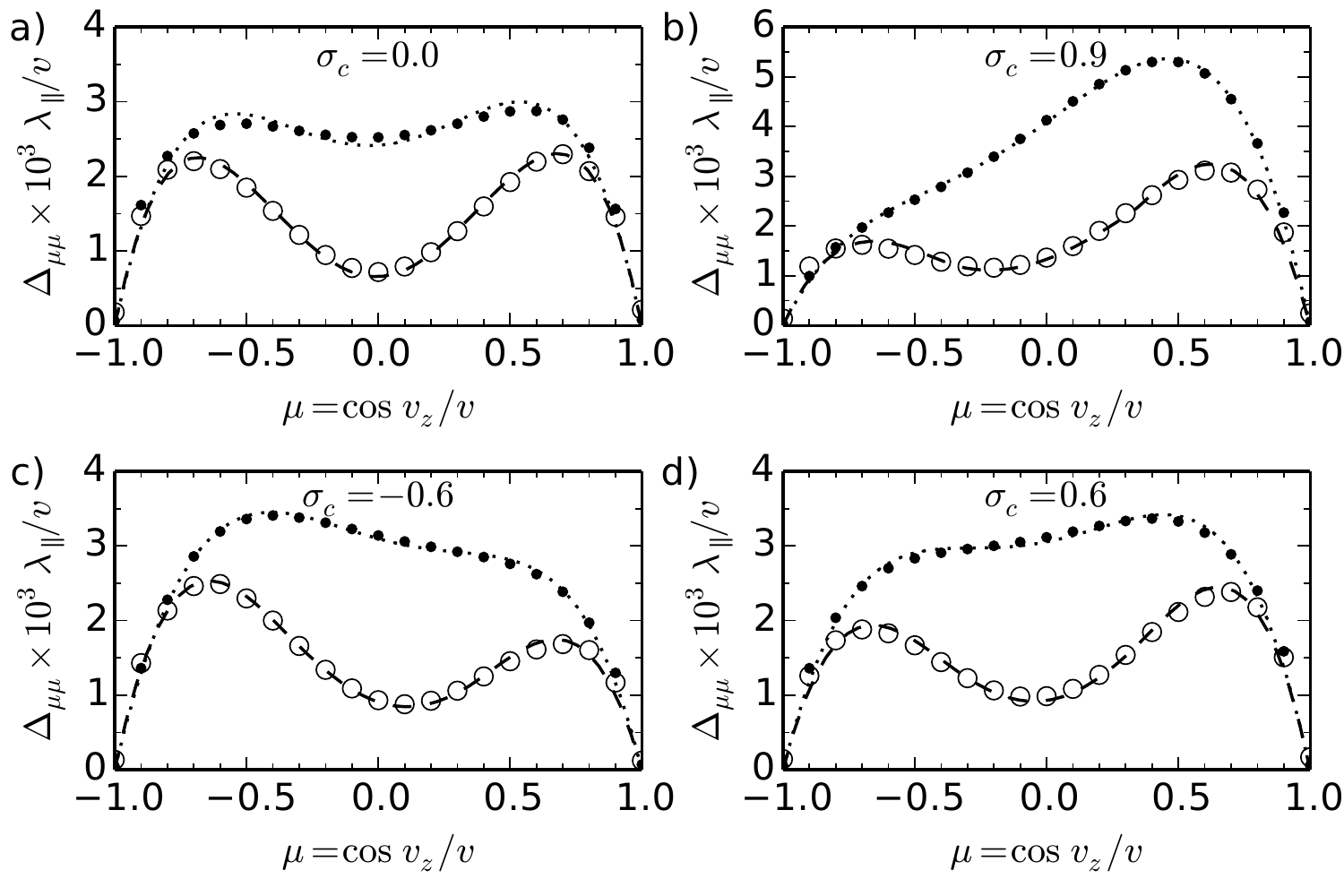}
    \caption{Magnetodynamic case: pitch-angle scattering rate $\Delmm(\mu)$ for cosmic-ray particles determined numerically after one gyroperiod and from a Levenberg-Marquardt fit to eqn.~(\ref{eqnDmm3}). See caption of figure~\ref{figED}}
    \label{figME}
\end{figure*}

In the `magnetodynamic' runs presented in this subsection, the MHD fields were still dynamically evolved in parallel with the test-particles. Since the acceleration is perpendicular to the velocity at each time step, the kinetic energy of each test-particle is perfectly conserved. As shown in Figure~\ref{figEMMme}, the evolution of both ensembles in the balanced case and of the positive-$\mu$ ensemble in the imbalanced case is similar to the fully electrodynamic setting presented above.

However, the averaged magnetic moment of the ensemble with $\mu(0)=-0.6$ visibly increases on longer timescales. Similarly, the distribution of $\Mmag$ continues to broaden for at least ten gyroperiods, as indicated by the red shaded region in Figure~\ref{figEMMme}c. Although the ensemble in Landau resonance with the dominant Alfv\'en-wave population cannot be decelerated anymore because electric acceleration is not considered, the coherent interaction with the magnetic component of the Alfv\'en waves is now able to violate the approximate conservation of the magnetic moment. Energy is transfered from the $v_z$-component into the perpendicular velocity components. Without a constant adiabatic invariant, the pitch-angle cosine can decrease as before, even though the kinetic energy is fixed, and $\Delta\mu^2(t)$ keeps increasing at a similar rate as in the fully electrodynamic runs. 

We use the first minima of $\Delta\mu^2$ for $\mu(0)\in\{-1,-0.9,\ldots,+1.0\}$ to determine a best-fit parameter set with the quasilinear $\Dmm$ in magnetodynamic turbulence for the same variables as before, but with $\Dmm$ given by equation~(\ref{eqnDmm3}). Most of the results are similar to those presented in the previous subsection, so we only show the best-fit cross-helicity values $\sigma\fit$ in Figure~\ref{figMEsig}. The results for the Alfv\'en-speed ensemble ($v(0)=v_A$) are highly inaccurate, while the ensemble with $v(0)=3v_A$ at least exhibits the correct trend: $\sigma\fit$ grows weakly as $\sigma$ increases.

Plotting $\Dmm$ and $\Delmm$ in Figure~\ref{figME}, we see that, without electric fields, the scattering asymmetry is strongly reduced compared to the graphs in Figure~\ref{figED}. Electric fields obviously play a crucial role in establishing this asymmetry in imbalanced turbulence.

\begin{figure}[ht]
\centering
\includegraphics{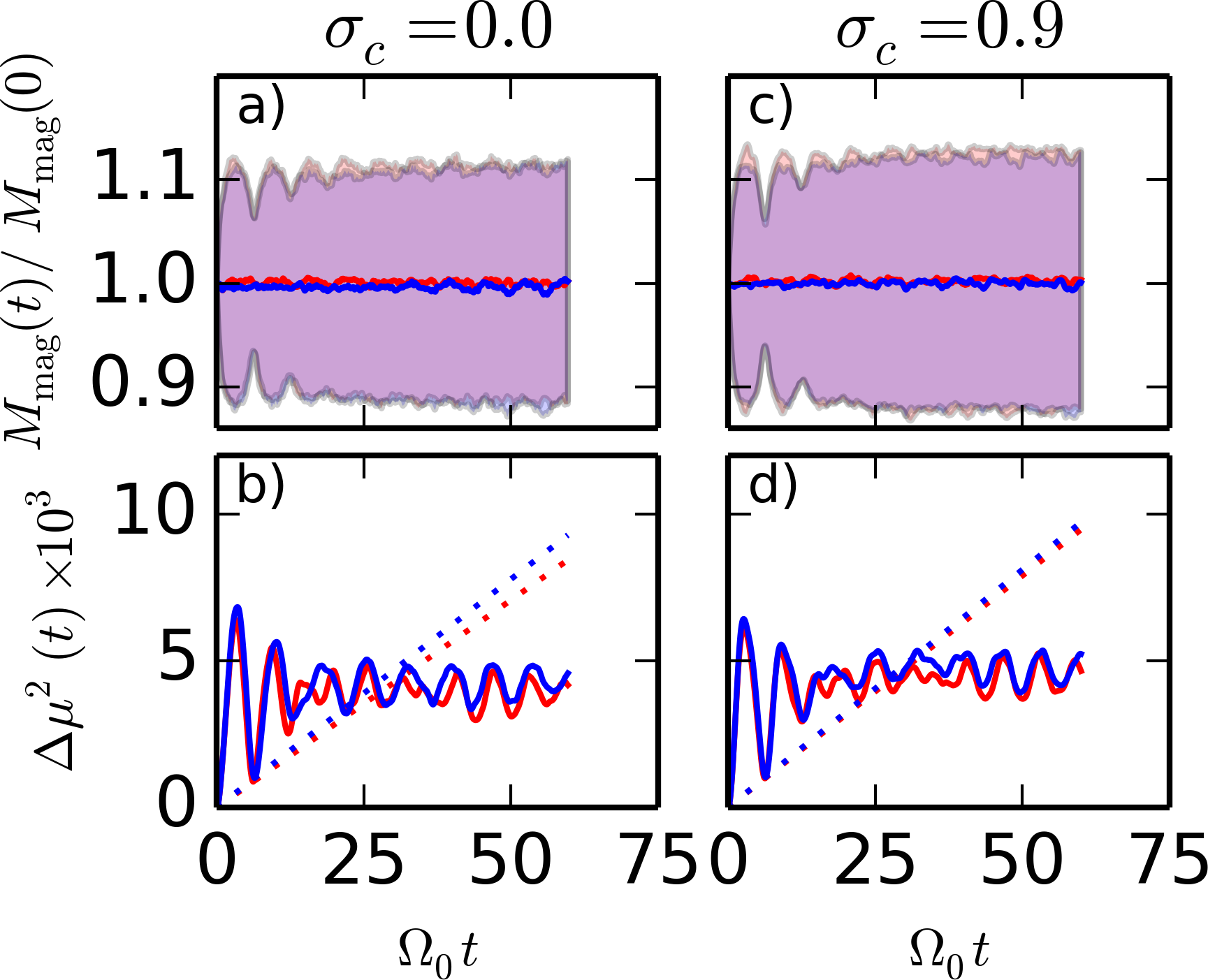}
\caption{Magnetostatic case: Evolution of $\Mmag$ and $\Delta\mu^2$ as in figure~\ref{figEMMbxu}}
\label{figEMMms}
\end{figure}

\begin{figure*}[ht]
    \centering
    \includegraphics{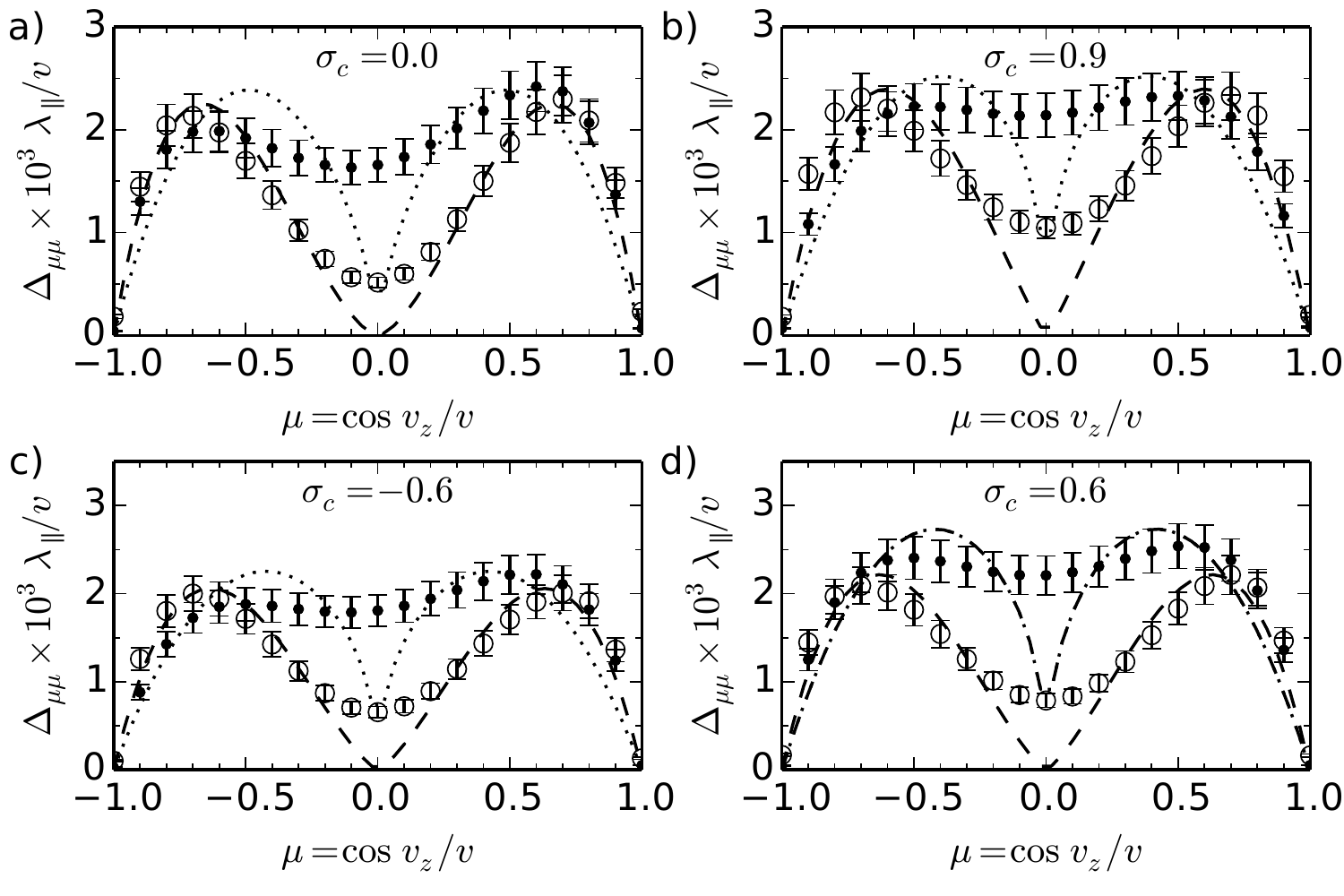}
    \caption{Magnetostatic case: pitch-angle scattering rate $\Delmm(\mu)$ for cosmic-ray particles determined numerically after one gyroperiod and from a Levenberg-Marquardt fit to eqn.~(\ref{eqnDmm1}). See caption of figure~\ref{figED}}
    \label{figMS}
\end{figure*}

\subsection{Magnetostatic runs}
\label{subMS}

To determine the relative importance of wave propagation, we have also performed test-particle simulations with single static snapshots of the magnetic-field configurations. In these magnetostatic runs, we calculate the Lorentz force on the particles using only the magnetic field, similar to the previous subsection (equation~(\ref{eqnLorentzMD})) but without evolving the field in time. 

Since particles can no longer interact coherently with traveling waves, Landau resonance as observed before is impossible and pitch-angle scattering proceeds at similar rates for positive and negative $\mu(0)$ even in strongly cross-helical turbulence. As shown in Figure~\ref{figEMMms}, the mean magnetic moment stays constant throughout the simulations and $\Delta\mu^2(t)$ settles on an approximately constant value after a few gyroperiods. 

Evolving the trajectories for longer times, which is possible since magnetostatic test-particle simulations are not very demanding compared to fully electrodynamic runs, yields a subdiffusive behavior of the pitch-angle distribution. Thus, we computed $\Delmm$ after one gyroperiod again to perform a best-fit operation, varying $\delta B$ and $s$ in equation~(\ref{eqnDmm1}). As we found statistical noise to be significantly larger in static runs than in the evolving-MHD runs presented above, we show a 10 \%-error bar in the plots of $\Delmm(\mu)$. Repeating each test-particle run in ten different static snapshots of the magnetic field for otherwise identical parameters, we found that the values we obtained for $\Delmm$ varied by this amount, whereas the evolving-MHD runs only varied by about 2 \%.

Taking this statistical noise into account, the graphs of $\Delmm(\mu)$ are symmetrical with respect to the sign of $\mu$ (see Figure~\ref{figMS}). Cross helicity is therefore irrelevant in magnetostatic turbulence, and the spatial structure of the magnetic fields can be ruled out as a contributor to the scattering asymmetry.  As a sidenote, we remark that none of our results were affected by the sign of the test-particle charge, proving that magnetic-helicity contributions to the scattering asymmetry, which would be sensitive to the particle charge, are insignificant.

\begin{figure*}[ht]
    \centering
    \includegraphics{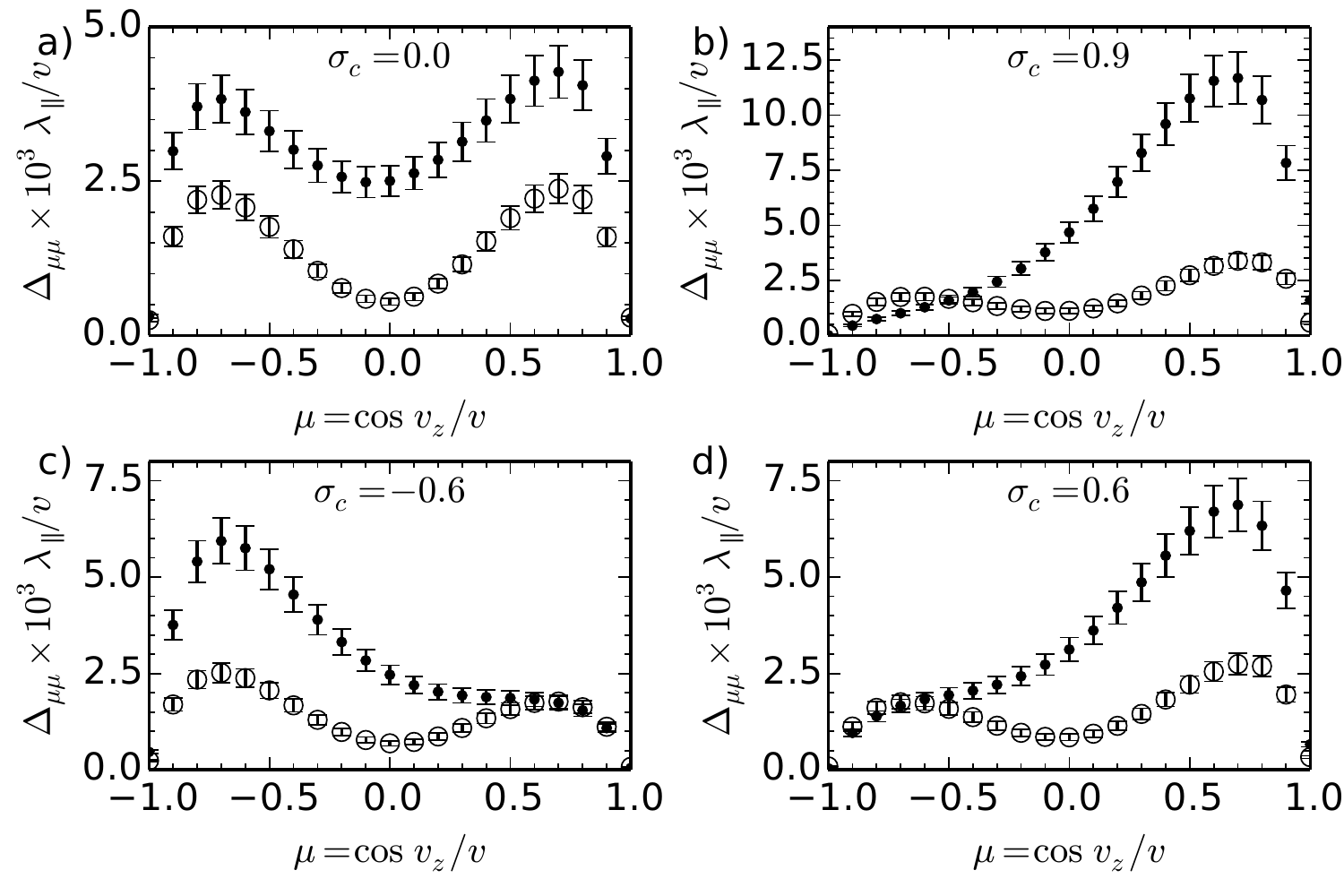}
    \caption{Static-electromagnetic case: pitch-angle scattering rate $\Delmm(\mu)$ for cosmic-ray particles determined numerically after one gyroperiod. See caption of figure~\ref{figED}}
    \label{figES}
\end{figure*}

\subsection{Static-electromagnetic runs}
\label{subES}

Adding the electric-field component back into the Lorentz force while still using only static field snapshots, we found that this asymmetry can be re-established (Figure~\ref{figES}). While the absolute scattering rates $\Delmm$ are reduced to about 70 \% compared to the values for the fully electrodynamic runs in Figure~\ref{figED}, the shapes of the graphs are almost identical to their fully electrodynamic equivalents in each case, especially if one takes the statistical noise into account. 

Since the electromagnetic fields of the Alfv\'en waves are fixed in time and space in these runs, it is the spatial structure of the electric fields that must be the predominant soure of the scattering asymmetry. Hence, we conclude that particles in fully evolving MHD turbulence with non-zero cross helicity are scattered differently for positive and negative pitch-angles not only because of coherent wave-particle interactions, but additionally because the spatial structure of electric fields in imbalanced turbulence favors such an asymmetry.
Of course, this structure can be viewed as the result of the imbalance between Alfv\'en-waves propagating parallel and anti-parallel to the magnetic mean-field direction, a property that is conserved when the fields are frozen.

\section{Conclusions}
\label{secConclusions}

We have investigated pitch-angle scattering of charged particles at velocities comparable to the Alfv\'en speed in realistically evolving magnetohydrodynamic turbulence with non-zero cross helicity. The dependence of $\Delmm$ on the initial pitch-angle is qualitatively well described by quasilinear diffusion theory under the assumption of a simple slab spectrum. The agreement with the theoretical prediction is particularly impressive in the case of fully electrodynamic turbulence, showing that, for example, the interaction of Alfv\'en-speed dust particles with shear-Alfv\'en waves they encounter in the solar wind can be modeled accurately by equation~(\ref{eqnDmm2}), at least on short timescales. An almost perfect fit of the quasilinear slab model to the numerical data is even possible in strongly cross-helical turbulence, as we have shown; therefore the scattering of charged particles in patches of strongly imbalanced turbulence, such as in the fast solar wind, is also described accurately by quasilinear theory. 

However, the degree of imbalance that we have to assume in order to obtain this perfect fit is lower than the imbalance which a direct analysis of the correlation of the turbulent MHD fields yields. Our results thus imply that quasilinear theory tends to overestimate the asymmetry of pitch-angle scattering in turbulence with extremely high cross helicity.

{As we noted in Section~\ref{secIntro}, pitch-angle scattering over small timescales, on which we have focused in this article, is important for theories describing the injection of relatively low-energy ions into the Fermi-acceleration mechanism at quasiperpendicular shocks \citep{kirk1989particle}. The structure of the shock-foot region is particularly sensitive to the short-time behaviour reflected in the shape of $\Delmm(\mu)$. Hence, our results will benefit future theoretical work on quasiperpendicular shocks. Although in such applications the overall magnetic turbulence will be stronger than the intermediate levels that we have considered here, these results can easily be transferred to pitch-angle scattering with regard to a magnetic mean-field averaged only over intermediate lengthscales. On such smaller scales, the anisotropy with respect to the mean-field is more pronounced \citep{howes2011gyrokinetic}, and our results are valid. Eventually, the local pitch-angle diffusion thus derived could be combined with models for fieldline diffusion and crossfield diffusion to obtain a self-consistent description of the shock-foot region, conceptually similar to compound diffusion \citep{kota2000velocity}.}

It should be noted that our simulations have focused on incompressible magnetohydrodynamics and did not include transit-time damping by the fast magnetosonic wave. The compressional fast mode has been shown to be more efficient at accelerating and scattering charged particles \citep{schlickeiser1998quasilinear,yan2002scattering} than the shear-Alfv\'en modes included in this investigation. However, the assumption of incompressible MHD and hence of a negligible contribution of the fast mode is still valid in many astrophysical systems, such as the fast solar wind.

In our magnetodynamic simulations, which differ from the previously described fully electrodynamic case in that we have neglected the acceleration force due to the electric fields, the pitch-angle asymmetry of the scattering during the first gyroperiod is significantly smaller. In static-electromagnetic runs, in which the electric force on the test-particles was included but the turbulent fields were frozen in time, this asymmetry is almost as pronounced as in the electrodynamic runs. This contrast implies that the pitch-angle dependence of the scattering coefficients in imbalanced turbulence is mainly due to the spatial structure of the electric fields, and only secondarily due to coherent wave-particle interactions, with possibly important consequences for the isotropization rate of cosmic-ray particles in shock fronts. It is clear from these observations that the predictive power of quasilinear theory for the diffusion of cosmic-ray particles in imbalanced turbulence profits immensely from using a more realistic spectral description of cross-helical fields.

On timescales exceeding two or three gyroperiods, we observed that the mean-square deviation of the pitch-angle cosine settled on an almost constant value for most test-particle ensembles, in contradiction to the diffusive broadening one would expect. This spreading continues for significantly longer if the test-particles are close to Landau resonance with the dominating Alfv\'en-wave population in imbalanced turbulence. While the presence of electric fields allows for an energy change that accompanies the longer-lasting spread of $\Delta\mu^2$, we found that magnetodynamic runs conserve energy perfectly, but violate the adiabatic invariance of the magnetic moment for such test-particles. Hence, without accounting for adiabatic focusing as described, for example, in \citet{schlickeiser2008cosmic} and \citet{schlickeiser2010cosmic}, the quasilinear description of pitch-angle scattering in imbalanced turbulence is valid, but only for a few gyroperiods.

M.S.W. would like to thank Nuno Loureiro for valuable discussions. This work was facilitated by the Max-Planck/Princeton Center for Plasma Physics. The research leading to the results presented in this article has received funding from the European Research Council under the European Union's Seventh Framework Programme (FP7/2007-2013)/ERC Grant Agreement No.~277870.

\appendix
\section{The one-period resonance function $\RR(T_g)$}
\label{secAppendix}

In standard applications of quasilinear theory, the resonance condition $\omega_R = k_z v_z \pm \Omega$ is obtained from the infinite-time limit of the resonance function $\RR(t)$. For example, the pitch-angle diffusion coefficient in the magnetostatic-turbulence model is
\begin{equation}
  \Dmm = \lim_{t\to\infty} \int \D k_z\:\frac{\Omega^2}{2}\ \frac{(1-\mu^2)}{|v \mu|^{2}}\ \frac{|\db_\perp(k_z)|^2}{B_0^2}\ \RR(t),
  \label{eqnQLTapp}
\end{equation}
where the resonance function is defined as
\begin{equation}
  \RR(t) = \Re \int_0^t \D\tau \exp(\I \varpi \tau - \Gamma \tau),
\end{equation}
with $\varpi = \omega_R - k_z v_z \pm \Omega$. Here $\Gamma$ denotes the damping rate of the oscillations. For general $t$, performing the integration and taking the real part yields
\begin{equation}
  \RR(t) = \frac{\Gamma\ [1-e^{-\Gamma\:t}\cos(\varpi t)] + \varpi\ e^{-\Gamma\:t} \sin(\varpi t)}{\Gamma^2 + \varpi^2},
\end{equation}
the infinite-time limit of which ($\Gamma t \gg 1$) we denote as $\RR_\infty$:
\begin{equation}
  \RR_\infty = \lim_{t\to\infty} \RR(t) = \frac{\Gamma}{\Gamma^2 + \varpi^2}.
\end{equation}

Thus, the resonance function can be written as
\begin{equation}
  \RR(t) = \RR_\infty + \RRcorr(t),
\end{equation}
where we have defined a correction term as
\begin{equation}
  \RRcorr(t) = e^{-\Gamma\:t}\ \frac{\varpi \sin(\varpi t) - \Gamma\ \cos(\varpi t)}{\Gamma^2 + \varpi^2}.
\end{equation}

In the weak-damping limit ($\Gamma \to 0$), $\RR_\infty$ reduces to a Dirac function,
\begin{equation}
  \lim_{\Gamma\to0} \RR_\infty = \pi\ \delta(\varpi),
\end{equation}
such that performing the integral in equation~(\ref{eqnQLTapp}) yields the standard result stated in equation~(\ref{eqnDmm1}). For finite times, however, $\RRcorr(t)$ gives an additional contribution even in the limit $\Gamma\to0$:
\begin{equation}
  \lim_{\Gamma\to0} \RR(t) = \pi\ \delta(\varpi) + \frac{\sin(\varpi t)}{\varpi}.
\end{equation}
Since the latter term is positive for $t=T_g$, the quasilinear result $\Dmm$ represents a lower limit for the pitch-angle scattering rate $\Delmm$ defined in equation~(\ref{eqnDelmm}).

\bibliographystyle{apj}
\bibliography{refs}

\begin{thebibliography}{}
\expandafter\ifx\csname natexlab\endcsname\relax\def\natexlab#1{#1}\fi

\bibitem[{{Arzner} {et~al.}(2006){Arzner}, {Knaepen}, {Carati}, {Denewet}, \&
  {Vlahos}}]{arzner2006effect}
{Arzner}, K., {Knaepen}, B., {Carati}, D., {Denewet}, N., \& {Vlahos}, L. 2006,
  \apj, 637, 322

\bibitem[{{Beresnyak} {et~al.}(2011){Beresnyak}, {Yan}, \&
  {Lazarian}}]{beresnyak2011numerical}
{Beresnyak}, A., {Yan}, H., \& {Lazarian}, A. 2011, \apj, 728, 60

\bibitem[{{Brandenburg} \& {Urpin}(1998)}]{brandenburg1998magnetic}
{Brandenburg}, A., \& {Urpin}, V. 1998, \aap, 332, L41

\bibitem[{{Breech} {et~al.}(2003){Breech}, {Matthaeus}, {Milano}, \&
  {Smith}}]{breech2003probability}
{Breech}, B., {Matthaeus}, W.~H., {Milano}, L.~J., \& {Smith}, C.~W. 2003,
  \jgr, 108, 1153

\bibitem[{{Dalena} {et~al.}(2012){Dalena}, {Greco}, {Rappazzo}, {Mace}, \&
  {Matthaeus}}]{dalena2012magnetic}
{Dalena}, S., {Greco}, A., {Rappazzo}, A.~F., {Mace}, R.~L., \& {Matthaeus},
  W.~H. 2012, \pre, 86, 016402

\bibitem[{{Dung} \& {Schlickeiser}(1990)}]{dung1990influence}
{Dung}, R., \& {Schlickeiser}, R. 1990, \aap, 240, 537

\bibitem[{{Goldstein}(1980)}]{goldstein1980mean}
{Goldstein}, M.~L. 1980, \jgr, 85, 3033

\bibitem[{{Grappin} {et~al.}(1983){Grappin}, {Leorat}, \&
  {Pouquet}}]{grappin1983dependence}
{Grappin}, R., {Leorat}, J., \& {Pouquet}, A. 1983, \aap, 126, 51

\bibitem[{{Hoang} {et~al.}(2012){Hoang}, {Lazarian}, \&
  {Schlickeiser}}]{hoang2012revisiting}
{Hoang}, T., {Lazarian}, A., \& {Schlickeiser}, R. 2012, \apj, 747, 54

\bibitem[{{Howes} {et~al.}(2011){Howes}, {Tenbarge}, {Dorland}, {Quataert},
  {Schekochihin}, {Numata}, \& {Tatsuno}}]{howes2011gyrokinetic}
{Howes}, G.~G., {Tenbarge}, J.~M., {Dorland}, W., {et~al.} 2011, Physical
  Review Letters, 107, 035004

\bibitem[{{Jaekel}(1998)}]{jaekel1998magnetic}
{Jaekel}, U. 1998, \pre, 58, 4033

\bibitem[{{Jokipii}(1966)}]{jokipii1966cosmic}
{Jokipii}, J.~R. 1966, \apj, 146, 480

\bibitem[{{Kirk} \& {Heavens}(1989)}]{kirk1989particle}
{Kirk}, J.~G., \& {Heavens}, A.~F. 1989, \mnras, 239, 995

\bibitem[{{K{\'o}ta} \& {Jokipii}(2000)}]{kota2000velocity}
{K{\'o}ta}, J., \& {Jokipii}, J.~R. 2000, \apj, 531, 1067

\bibitem[{Landau \& Lifshitz(1982)}]{landau1982mechanics}
Landau, L., \& Lifshitz, E. 1982, Mechanics No.~1 (Elsevier Science)

\bibitem[{{Lazarian} \& {Yan}(2002)}]{lazarian2002grain}
{Lazarian}, A., \& {Yan}, H. 2002, \apjl, 566, L105

\bibitem[{{Lehe} {et~al.}(2009){Lehe}, {Parrish}, \&
  {Quataert}}]{lehe2009heating}
{Lehe}, R., {Parrish}, I.~J., \& {Quataert}, E. 2009, \apj, 707, 404

\bibitem[{{Lynn} {et~al.}(2012){Lynn}, {Parrish}, {Quataert}, \&
  {Chandran}}]{lynn2012resonance}
{Lynn}, J.~W., {Parrish}, I.~J., {Quataert}, E., \& {Chandran}, B.~D.~G. 2012,
  \apj, 758, 78

\bibitem[{{Marsch} \& {Tu}(1990)}]{marsch1990radial}
{Marsch}, E., \& {Tu}, C.-Y. 1990, \jgr, 95, 8211

\bibitem[{Matthaeus {et~al.}(2004)Matthaeus, Minnie, Breech, Parhi, Bieber, \&
  Oughton}]{matthaeus2004transport}
Matthaeus, W.~H., Minnie, J., Breech, B., {et~al.} 2004, \grl, 31, L12803

\bibitem[{{Michalek} \& {Ostrowski}(1998)}]{michalek1998cosmic}
{Michalek}, G., \& {Ostrowski}, M. 1998, \aap, 337, 558

\bibitem[{Perez \& Boldyrev(2009)}]{perez2009role}
Perez, J.~C., \& Boldyrev, S. 2009, \prl, 102, 025003

\bibitem[{{Podesta}(2009)}]{podesta2009dependence}
{Podesta}, J.~J. 2009, \apj, 698, 986

\bibitem[{{Qin} \& {Shalchi}(2009)}]{qin2009pitch}
{Qin}, G., \& {Shalchi}, A. 2009, \apj, 707, 61

\bibitem[{{Sagdeev}(1966)}]{sagdeev1966}
{Sagdeev}, R.~Z. 1966, Reviews of Plasma Physics, 4, 23

\bibitem[{{Schlickeiser}(1989)}]{schlickeiser1989cosmic}
{Schlickeiser}, R. 1989, \apj, 336, 243

\bibitem[{Schlickeiser(2002)}]{schlickeiser2002cosmic}
Schlickeiser, R. 2002, Cosmic Ray Astrophysics, Astronomy and Astrophysics
  Library (Springer)

\bibitem[{{Schlickeiser} \& {Jenko}(2010)}]{schlickeiser2010cosmic}
{Schlickeiser}, R., \& {Jenko}, F. 2010, Journal of Plasma Physics, 76, 317

\bibitem[{{Schlickeiser} \& {Miller}(1998)}]{schlickeiser1998quasilinear}
{Schlickeiser}, R., \& {Miller}, J.~A. 1998, \apj, 492, 352

\bibitem[{{Schlickeiser} \& {Shalchi}(2008)}]{schlickeiser2008cosmic}
{Schlickeiser}, R., \& {Shalchi}, A. 2008, \apj, 686, 292

\bibitem[{{Shalchi}(2009)}]{shalchi2009diffusive}
{Shalchi}, A. 2009, Astroparticle Physics, 31, 237

\bibitem[{Shalchi(2009)}]{shalchi2009nonlinear}
Shalchi, A. 2009, Nonlinear Cosmic Ray Diffusion Theories, Astrophysics and
  Space Science Library (Springer)

\bibitem[{{Shalchi} {et~al.}(2009){Shalchi}, {Koda}, {Tautz}, \&
  {Schlickeiser}}]{shalchi2009analytical}
{Shalchi}, A., {Koda}, T.~{\AA}., {Tautz}, R.~C., \& {Schlickeiser}, R. 2009,
  \aap, 507, 589

\bibitem[{{Tautz} {et~al.}(2013){Tautz}, {Dosch}, {Effenberger}, {Fichtner}, \&
  {Kopp}}]{tautz2013pitch}
{Tautz}, R.~C., {Dosch}, A., {Effenberger}, F., {Fichtner}, H., \& {Kopp}, A.
  2013, \aap, 558, A147

\bibitem[{{Teaca} {et~al.}(2009){Teaca}, {Verma}, {Knaepen}, \&
  {Carati}}]{teaca2009energy}
{Teaca}, B., {Verma}, M.~K., {Knaepen}, B., \& {Carati}, D. 2009, \pre, 79,
  046312

\bibitem[{{Teaca} {et~al.}(2014){Teaca}, {Weidl}, {Jenko}, \&
  {Schlickeiser}}]{teaca2014acceleration}
{Teaca}, B., {Weidl}, M.~S., {Jenko}, F., \& {Schlickeiser}, R. 2014, \pre, 90

\bibitem[{{Yan} \& {Lazarian}(2002)}]{yan2002scattering}
{Yan}, H., \& {Lazarian}, A. 2002, \prl, 89, 1102

\bibitem[{{Yoshizawa} \& {Yokoi}(1993)}]{yoshizawa1993turbulent}
{Yoshizawa}, A., \& {Yokoi}, N. 1993, \apj, 407, 540

\end{thebibliography}

\end{document}